\documentclass[11pt]{article}
\usepackage{marvosym, amsfonts, floatflt,graphicx, multicol, color,cite}

\usepackage{ifpdf}
\ifpdf
\usepackage{hyperref, pdfsync, epstopdf}	
\else
\usepackage[dvips,bookmarks]{hyperref}	
\fi
\hypersetup{colorlinks,bookmarksopen,bookmarksnumbered,citecolor=verdes,
linkcolor=blus,pdfstartview=FitH,urlcolor=rossos}
\def\hhref#1{\href{http://arxiv.org/abs/#1}{#1}} 

\def\be{\begin{equation}}
\def\ee{\end{equation}}

\newcommand{\beq}{\begin{equation}}
\newcommand{\eeq}{\end{equation}}
\newcommand{\bea}{\begin{eqnarray}}
\newcommand{\eea}{\end{eqnarray}}
\newcommand{\SU}{\hbox{SU}}
\newcommand{\fig}[1]{~{\rm \ref{fig:#1}}}

\def\circa#1{\,\raise.3ex\hbox{$#1$\kern-.75em\lower1ex\hbox{$\sim$}}\,}
\newcommand{\ds}{\partial\hspace{-1.3ex}/}
\newcommand{\eV}{\,{\rm eV}}

\newcommand{\MeV}{\,{\rm MeV}}
\newcommand{\GeV}{\,{\rm GeV}}
\newcommand{\TeV}{\,{\rm TeV}}
\newcommand{\eq}[1]{~(\ref{eq:#1})} 
\font\tenrsfs=rsfs10
\font\sevenrsfs=rsfs7
\font\fiversfs=rsfs5
\newfam\rsfsfam
\textfont\rsfsfam=\tenrsfs
\scriptfont\rsfsfam=\sevenrsfs
\scriptscriptfont\rsfsfam=\fiversfs
\def\mathscr#1{{\fam\rsfsfam\relax#1}}
\def\Lag{\mathscr{L}}

\newcommand{\mb}[1]{\mbox{\normalsize\boldmath $#1$}}

\newcommand{\epsN}{\varepsilon_1}
\newcommand{\AL}{A_L}
\newcommand{\YBL}{A_{B-L}}
\newcommand{\YB}{A_B}
\newcommand{\YL}{A_L}

\renewcommand{\AL}{{\cal L}}
\renewcommand{\YBL}{{\cal B}-{\cal L}}
\renewcommand{\YB}{{\cal B}}
\renewcommand{\YL}{{\cal L}}

\makeatletter
\def\sottonumero{}
\def\art{\@ifnextchar[{\eart}{\oart}}
\def\eart[#1]#2#3#4#5#6{{\rm #2}, {\em #3 \rm #4} {\rm (#6) #5 ({\em\hhref{#1}})}\sottonumero}
\def\hepart[#1]#2{{\rm #2, \em\hhref{#1}}\sottonumero}
\newcommand{\oart}[5]{{\rm #1}, {\em #2 \rm #3} {\rm (#5) #4}\sottonumero}

\newcommand{\NP}{Nucl. Phys.}

\newcommand{\PRL}{Phys. Rev. Lett.}
\newcommand{\PL}{Phys. Lett.}
\newcommand{\PR}{Phys. Rev.}

\definecolor{rosso}{cmyk}{0,1,1,0.4}
\definecolor{rossos}{cmyk}{0,1,1,0.55}
\definecolor{rossoc}{cmyk}{0,1,1,0.2}
\definecolor{blu}{cmyk}{1,1,0,0.3}
\definecolor{blus}{cmyk}{1,1,0,0.6}
\definecolor{bluc}{cmyk}{1,1,0,0.1}
\definecolor{verde}{cmyk}{0.92,0,0.59,0.25}
\definecolor{verdec}{cmyk}{0.92,0,0.59,0.15}
\definecolor{verdes}{cmyk}{0.92,0,0.59,0.4}
\definecolor{giallo}{cmyk}{0,0,1,0}
\definecolor{gialloverde}{cmyk}{0.44,0,0.74,0}

\newcommand{\mio}[1]{}

\begin{document}
\begin{center}
\color{black}
{\Huge\bf\color{rossos}
Baryogenesis via leptogenesis}
\medskip
\bigskip\color{black}\vspace{0.5cm}

{
{\large\bf Alessandro Strumia}
}
\\[7mm]
{\it Dipartimento di Fisica dell'Universit{\`a} di Pisa and INFN, Italia}\\
\end{center}

\bigskip

\centerline{\large\bf\color{blus} Abstract}
\begin{quote}\large\color{blus}
We discuss how leptogenesis can explain the observed baryon asymmetry
and summarize attempts of testing leptogenesis.
We first perform estimates and discuss the main physics, and later 
outline the techniques that allow to perform precise computations.

\color{black}
\end{quote}

\vfill

\tableofcontents

\newpage

\section{Introduction}\label{Introduction} 
The universe contains various relict particles: 
$\gamma$,  $e$, $p$, $\nu$, $^4{\rm He}$, Deuterium,\ldots, 
plus likely some  Dark Matter (DM).
Their abundances are mostly understood,
with the following main exception:
\begin{equation}
\label{eq:nB}
 \frac{n_B - n_{\bar B}}{n_\gamma} = 
\frac{n_B}{n_\gamma} =(6.15\pm0.25)\cdot 10^{-10}
\end{equation}
where $n_\gamma$  and $n_B$ and are the present number densities of photons  and baryons 
(anti-baryons have negligible density).
This is the problem of baryogenesis.
Before addressing it, let us briefly summarize the analogous understood issues.

As suggested by inflation, the total energy density equals the critical energy density,
discussed later.
Next, almost all relative abundances can be understood
assuming that these particles are thermal relics of a hot Big-Bang phase.
The number densities of electrons and protons are equal,
$n_e = n_p$, because nothing violated electric charge.
The relative proton/neutron abundancy was fixed by electroweak processes such as
$n\nu_e\leftrightarrow pe$ at $T\sim {\rm few~MeV}$.
Neutrons get bound in nuclei at  $T\sim 0.1\MeV$:
the measured  nuclear primordial nuclear abundances  agree with predictions: 
$n_{^4{\rm He}}/n_p\approx 0.25/4$,
$n_{\rm D}/n_p\approx 3~10^{-5}/2$, etc.
The neutrino density, predicted to be $n_{\nu_{e,\mu,\tau}} =n_{\bar\nu_{e,\mu,\tau}}= 3n_\gamma/22$, 
is too low to be experimentally tested:
the baryon asymmetry problem is more pressing than the analogous
 lepton asymmetry problem because we do not know how to measure the lepton asymmetry.
Finally, the DM abundancy suggested by present data 
is obtained if DM particles are weakly-interacting thermal relicts with mass 
\beq\label{eq:mDM}
 m \sim \sqrt{T_{\rm now}\cdot M_{\rm Pl}}\sim\TeV\eeq 
where $T_{\rm now}\approx 3\,{\rm K}$ is the present temperature
and $M_{\rm Pl}\approx 1.2~10^{19}\GeV$ is the Planck mass.
The LHC collider might soon produce DM particles and test this speculation.
It is useful to digress and understand eq.\eq{mDM},
because the necessary tools will reappear, in
a more complicated context, when discussing leptogenesis.

\subsection{The DM abundancy}\label{DM}
First, we need to compute the expansion rate $H(t)$ of the universe
as function of its energy density $\rho(t)$.
Let us study how a homogeneous  $\rho(t)$ made of non-relativistic matter
evolves according to gravity.
A test particle at distance $R$ from us feels the Newton acceleration
\beq \label{eq:ddotR}
\ddot{R}= - \frac{G M(R)}{R^2}=
-\frac{4\pi G\rho(t)}{3} R \eeq
where $M(R)$ is the total mass inside a sphere of radius $R$ and
$G=1/M_{\rm Pl}^2$ is the Newton constant.
By multiplying both sides of eq.\eq{ddotR} times $\dot{R}$ and integrating 
taking into account that   $\rho(t)\propto 1/R^3(t)$ one obtains
as usual the `total energy' constant of motion, here named $k$:
\beq\label{eq:H}\frac{d}{dt}\left[ \frac{1}{2}\dot{R}^2-\frac{4\pi}{3}G\rho R^2\right]=0\qquad\hbox{so that}\qquad
{ H^2\equiv \frac{\dot{R}^2}{R^2} = \frac{8\pi G}{3}\rho} - \frac{k}{R^2}.\eeq
Let us discuss the special case $k=0$.
It is obtained when the density $\rho$ equals the `critical density'
$\rho = \rho_{\rm cr}\equiv 3H^2/8\pi G$.
$k=0$ is special because means zero `total energy'
(the negative gravitational potential energy compensates the positive matter energy):
a universe with critical density that expands getting big for free
could have been theoretically anticipated since 1687.
Today we abandoned prejudices for a static universe, and
more advanced theories put the above discussion on firmer grounds.
In general relativity eq.\eq{H} holds for more general sources of energy
(relativistic particles, cosmological constant,...), and the constant $k$ gets the physical
meaning of curvature of the universe.
The inflation mechanism generates a smooth universe with negligibly small $k$.

\bigskip

Second, we need to know that a gas of particles in thermal equilibrium at temperature $T\gg m$
has number density $n_{\rm eq} \sim T^3$ and energy density
$\rho_{\rm eq} \sim T^4$:
one particle with energy $\sim T$ per de-Broglie wavelength $\sim1/T$.
The number density of non relativistic particles ($T\ll m$) is suppressed by a
Boltzmann factor, $n_{\rm eq}\sim (mT)^{3/2} e^{-m/T}$, and their energy density is $\rho_{\rm eq}\simeq m n_{\rm eq}$.

\medskip

We can now understand eq.\eq{mDM},
by studying what happens to a DM particle of mass $m$ when the temperature $T$
cools below $m$.

\smallskip

\noindent
\parbox{0.6\textwidth}{

\indent
Annihilations  with cross section $\sigma$(DM DM $\to$ SM particles)
try to maintain
thermal equilibrium,
$n_{\rm DM}\propto \exp(-m/T)$.
But they fail at  $T\circa{<} m$, when $n_{\rm DM}$ is so small that
the collision rate $\Gamma$ experienced by a DM particle
 becomes smaller than the expansion rate $H$:
$$\Gamma \sim n_{\rm DM} \sigma ~ \circa{<} ~H \sim T^2/M_{\rm Pl}.$$
As illustrated in the picture,
annihilations become ineffective, leaving the following out-of-equilibrium relic abundancy
of DM particles:}
{\hspace{0.05\textwidth}\raisebox{-2.3cm}{~~\includegraphics[width=0.3\textwidth,height=0.3\textwidth]{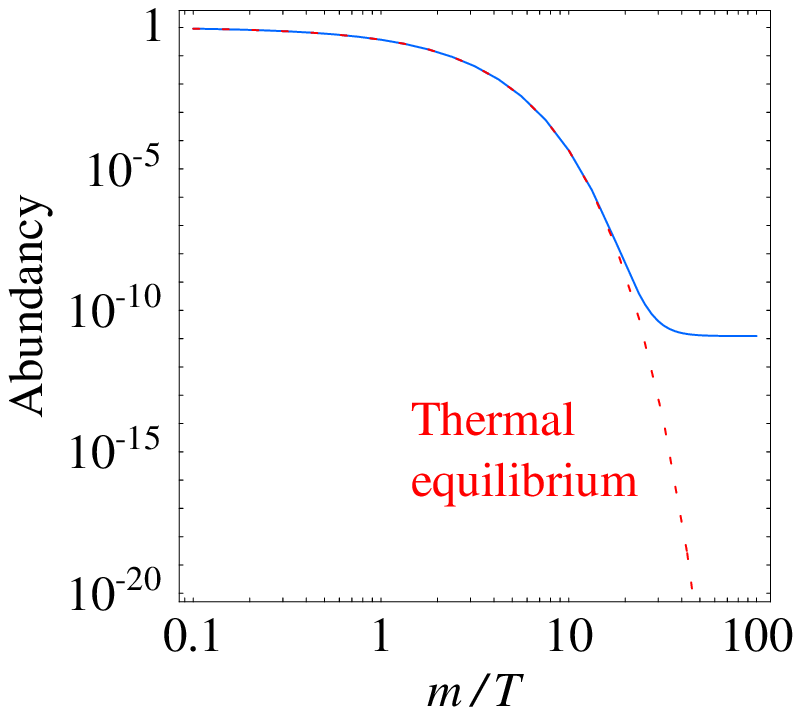}}}

\beq \label{eq:relic}\hspace{-4mm}\frac{n_{\rm DM}}{n_\gamma}
 \sim \frac{m^2/M_{\rm Pl}\sigma}{m^3}\sim
\frac{1}{M_{\rm Pl} \sigma m}\qquad\hbox{i.e.}\qquad
 \frac{\rho_{\rm DM}(T)}{\rho_{\gamma}(T)}\sim \frac{m}{T}
\frac{n_{\rm DM}}{n_{\gamma}} \sim \frac{1}{M_{\rm Pl}\sigma T}.\eeq
Inserting  the observed DM density, $\rho_{\rm DM} \sim \rho_\gamma$ at $T\sim T_{\rm now}$,
and a typical cross section
$\sigma\sim g^4/m^2$ gives eq.\eq{mDM} for a DM particle with weak coupling $g\sim 1$.
A precise computation can be done solving Boltzmann equations for DM.

\begin{figure}
$$\includegraphics[width=5cm]{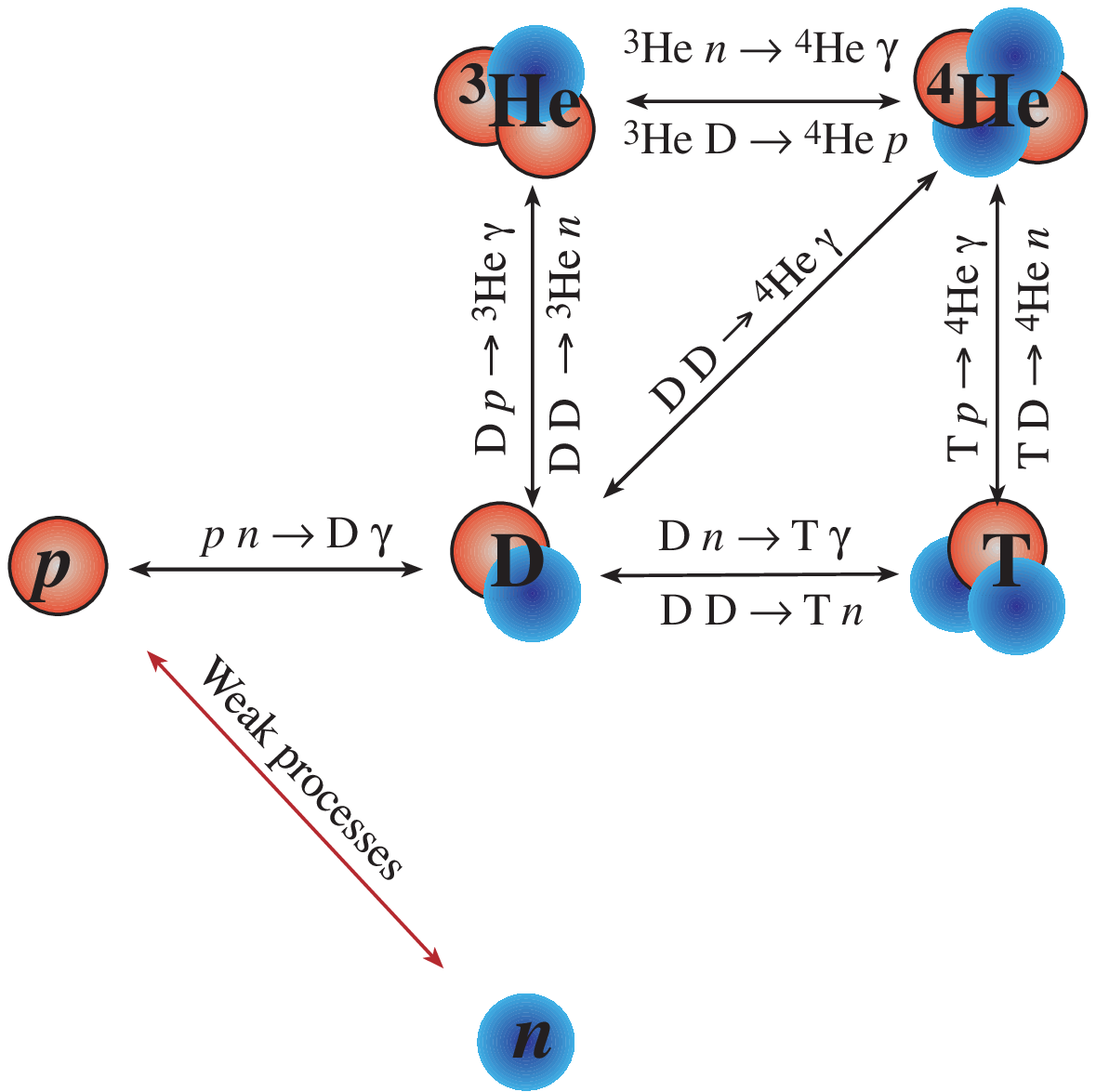}\qquad\raisebox{-5mm}{
 \includegraphics[height=5.5cm]{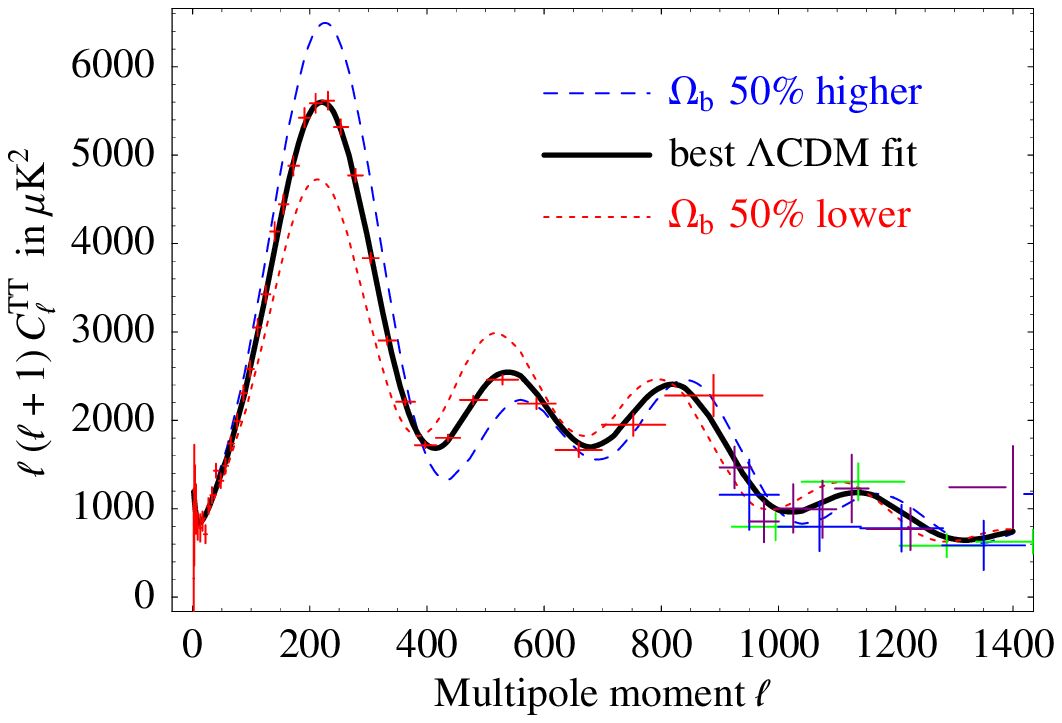}}$$
 \caption{\label{fig:exp}\em (a) Main reactions that determine primordial nuclear abundances.
 (b) How CMB anisotropies depend on the baryon abundancy $\Omega_B = \rho_B/\rho_{\rm cr}$,
 compared with data.}
\end{figure}

\subsection{The baryon asymmetry}
Let us summarize how the value of the baryon asymmetry in eq.\eq{nB}
is measured.
The photon density directly follows from the measurement
of the CMB  temperature  and from Bose-Einstein statistics:
$n_\gamma \sim T^3$.
Counting baryons is more difficult.
Direct measurements are not accurate,
because only some fraction of baryon formed stars and other luminous objects.
Two different indirect probes point to the same baryon density, making the result trustable.
Each one of the two probes would require a dedicated lesson:
\begin{itemize}
\item[1.] Big-Bang-Nucleosynthesis (BBN) predictions depend on $n_B/n_\gamma$.
Fig.\fig{exp}a illustrates the main reactions.
The important point is that the  presence of many more photons than baryons
delays BBN, mainly by enhancing the reaction
$pn\leftrightarrow \hbox{D}\gamma$ 
in the $\leftarrow$ direction, so that D forms not when the
temperature equals the Deuterium binding energy $B\approx 2\MeV$,
but later at $T\approx B /\ln(n_B/n_\gamma)\approx 0.1\MeV$,
giving more time to free neutrons to decay.
A precise computation can be done solving Boltzmann equations for neutrino
decoupling and nucleosynthesis.

\item[2.]
Measurements of CMB anisotropies~\cite{WMAP},
among many other things, allow us to probe 
acoustic oscillations of the baryon/photon fluid
happened around photon last scattering.
A precise computation can be done evolving Boltzmann equations for  anisotropies,
assuming that they are generated by quantum fluctuations during inflation:
fig.\fig{exp}b illustrates how the amount of anisotropies with angular scale $\sim 1/\ell$ 
depends on $n_B/n_\gamma$.
Acoustic oscillations have been seen also in matter inhomogeneities~\cite{WMAP},
at $\approx 3\sigma$ level.

\end{itemize}

\section{Baryogenesis}
The small baryon asymmetry $n_B/n_\gamma \ll 1$ 
can be obtained from a hot big-bang as the result of a
small excess of baryons over anti-baryons.
We would like to understand why,
when at $T\circa{<}m_p$ matter  almost completely annihilated with anti-matter,
we survived thanks to the `almost':
$$n_B - n_{\bar B}\propto 1000000001-1000000000 = 1.$$
This might be the initial condition at the beginning of the big-bang,
but it would be a surprisingly small excess.
In inflationary models it is regarded as a surprisingly large excess,
since inflation erases initial conditions.

In absence of a baryon asymmetry an equal number of relic baryons and of anti-baryons
survive to annihilations at $T\circa{<}m_p$. This process is analogous to DM annihilations
studied in the previous section, so we can  estimate  the relic
baryon density by inserting $m = m_p$ and a typical $p\bar{p}$ cross section $\sigma \sim 1/m_p^2$
in eq.\eq{relic}, obtaining $n_p/n_\gamma \sim m_p/M_{\rm Pl} \sim 10^{-19}$.
Therefore this is a negligible contribution.

Assuming that the hot-big-bang started with zero baryon asymmetry
at some temperature $T\gg m_p$, can the baryon asymmetry
can be generated dynamically in the subsequent evolution?
Once that one realizes that this is an interesting issue (this was done by Sakharov), 
the answer is almost obvious:   yes,  provided that at some stage~\cite{Baryogenesis}
\begin{itemize}
\item[1.] baryon number $B$ is violated; 
\item[2.]  C and CP are violated
(otherwise baryons and antibaryons behave in the same way);
\item[3.]  the universe was not in thermal equilibrium
(we believe that CPT is conserved,  so that particles and
antiparticles have the same mass, 
and therefore in thermal equilibrium have the same abundance).
\end{itemize}
Having discussed in section~\ref{DM} a concrete out-of-equilibrium situation,
it should be clear what the general concept means.

\medskip

\subsection{Baryogenesis in the SM?}
A large amount of theoretical and experimental work showed that,
{\em within the SM, Sakharov conditions are not fulfilled}.
At first sight one might guess that the only problem is 1.;
in reality 2.\ and 3.\ are problematic.

\smallskip 

\begin{itemize}
\item[1.] Within the SM $B$ is violated in a non trivial way~\cite{Baryogenesis},
thanks to quantum anomalies combined with extended $\SU(2)_L$ field-configurations:
the anomalous $B$ and $L$ symmetry are violated, while
$B-L$ is a conserved anomaly-free symmetry.
The basic equation is $\partial_\mu J^\mu_B \sim N_{\rm gen} F_{\mu\nu}^a\tilde{F}_{\mu\nu}^a$,
and implies that there are many vacua that differ by their $B,L$ content, separated
by a potential barrier of electroweak height.
At temperatures $T \ll 100\GeV$ transition between different vacua are negligible because
suppressed by a quantum-tunneling  factor 
$ e^{-2\pi /\alpha_2}$.
If $N_{\rm gen}=1$ this would imply proton decay with an unobservably slow rate;
since there are 3 generations and all of them must be involved, proton decay is kinematically forbidden.
This suppression disappears at high temperature, $T\circa{>}100\GeV$, and the 
space-time density of $B,L$-violating `sphaleron' interactions is $\gamma\sim \alpha_2^5 T^4$, 
faster than the expansion rate of the universe up to temperatures of about $T\sim 10^{12}\GeV$~\cite{Baryogenesis}.
 \footnote{A real understanding of these issues needs  advanced quantum field theory.
This kind of theoretical studies lead to one observed consequence:
the $\eta'$ mass, that is related to some QCD analogous of the $\SU(2)_L$ effects we are considering.
Therefore there should be no doubt that $B,L$ are violated, and
this is almost all what one needs to know to understand leptogenesis quantitatively.}

 \item[3.]
SM baryogenesis is not possible due to the lack of out-of equilibrium conditions.
The electroweak phase transition was regarded as a potential out-of equilibrium stage, but 
experiments now demand a higgs mass $m_h \circa{>}115\GeV$,
and SM computations of the Higgs thermal potential show that, for $m_h \circa{>} 70\GeV$, 
the higgs vev shifts smoothly from $\langle h\rangle = 0$ to
$\langle h\rangle = v$ when the universe cools down below $T\sim m_h$~\cite{Baryogenesis}.

\item[2.]
In any case,
the amount of CP violation provided by the CKM phase would have been too small for generating the
observed baryon asymmetry, because it is suppressed by small quark masses.
Indeed CP-violation would be absent if the light quarks were massless.
 \end{itemize}

Many extensions of the SM could generate the observed $n_B$.
`Baryogenesis at the electroweak phase transition'
needs new particles coupled to the higgs 
in order to obtain a out-of-equilibrium phase transition
and to provide extra sources of CP violation.
This already disfavored possibility will be tested at future accelerators.
`Baryogenesis from decays of GUT particles'
seems to conflict with non-observation of magnetic monopoles, at least in simplest models.
Furthermore minimal GUT model do not violate $B-L$,
so that sphaleron processes would later wash out the eventually generated baryon asymmetry.

\medskip

The existence of sphalerons suggests {\em baryogenesis through leptogenesis}:
lepton number might be violated by some non SM physics,
giving rise to a lepton asymmetry, which is converted
into the observed baryon asymmetry by sphalerons.

This scenario can be realized in many different ways~\cite{Baryogenesis}.
Majorana neutrino masses violate lepton number and presumably CP,
but do not provide enough out-of-equilibrium processes.
The minimal successful implementation~\cite{leptogenesis} needs just the minimal
amount of new physics which can give the observed small neutrino masses
via the see-saw mechanism~\cite{seesaw}:
heavy right-handed neutrinos $N$ with masses $M$.
{\em `Baryogenesis via thermal  leptogenesis'}~\cite{leptogenesis} proceeds at $T\sim M$, when
out-of-equilibrium  (condition 3) CP-violating (condition 2) decays of heavy right-handed neutrinos
generate a lepton asymmetry, converted in baryon asymmetry by SM sphalerons (condition 1).


\begin{figure}
$$\includegraphics[width=\textwidth]{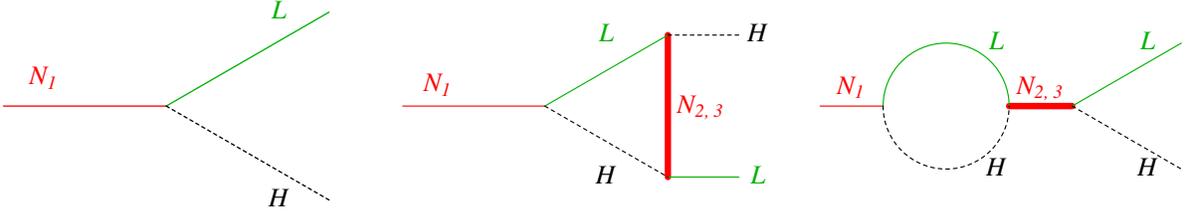}$$
\caption{\em CP-violating $N_1$ decay.\label{fig:Ndecay}}
\end{figure}

\section{Thermal leptogenesis: the basic physics}\label{lepto}
We now discuss the basic physics, obtaining estimates for the main results.
The SM is extended by adding the heavy right-handed neutrinos suggested by see-saw models.
To get the essential points, we consider a simplified model with one lepton doublet $L$
and two right-handed neutrinos,
that we name `$N_1$' and `$N_{2,3}$', with 
$N_1$ lighter than $N_{2,3}$.
The relevant Lagrangian is
\begin{eqnarray}\label{eq:Lseesaw}
\Lag &=& \Lag_{\rm SM}+ \bar N_{1} i \ds\, N_{1} + \lambda_1 N_1 HL+
 \frac{M_1}{2} N_1^2 +\\ \nonumber
 &&+ \bar N_{2,3} i \ds\, N_{2,3} + \lambda_{2,3} N_{2,3} HL + \frac{M_{2,3}}{2} N^{2}_{2,3}+\hbox{h.c.}
 \end{eqnarray}
By redefining the phases of the $N_1$, $N_{2,3}$, $L$ fields one can set $M_1$, $M_{2,3}$, $\lambda_1$ real
leaving an ineliminabile CP-violating phase in $\lambda_{2,3}$.


\smallskip

\subsection{CP-asymmetry}
The tree-level decay width of $N_1$ is  $ \Gamma_1 = {\lambda^2_1 M_1}/{8\pi} $.
The interference between tree and loop diagrams shown in fig.\fig{Ndecay}
renders $N_1$ decays CP-asymmetric:
$$  \epsN \equiv \frac{\Gamma(N_1\to L H)-\Gamma(N_1\to \bar{L} \bar H)}{\Gamma(N_1\to L H)+\Gamma(N_1\to \bar{L} \bar H)}\sim
\frac{1}{4\pi}\frac{M_1}{M_{2,3}}
{\rm Im}\,\lambda^{ 2}_{2,3}$$
In fact 
$$ \Gamma(N_1\to L H) \propto |\lambda_1 + A \lambda^*_1 \lambda^{2}_{2,3}|^2,\qquad
\Gamma(N_1\to \bar{L} \bar H) \propto |\lambda^*_1 + A \lambda_1 \lambda^{2* }_{2,3}|^2$$
where $A$ is the complex CP-conserving loop factor.
In the limit $M_{2,3}\gg M_1$ the sum of the two one loop diagrams
reduces to an insertion of the $(LH)^2$ neutrino mass operator mediated by $N_{2,3}$:
therefore $A$ is suppressed by one power of $M_{2,3}$.
The intermediate states in the loop diagrams in fig.\fig{Ndecay} can be on shell;
therefore the Cutkosky rule guarantees that $A$ has an imaginary part.
Inserting the numerical factor valid in the limit $M_{2,3}\gg M_1$
we can rewrite the CP-asymmetry as
\begin{equation}
\label{eq:epsilon1}{
\epsN \simeq  \frac{3}{16\pi} \frac{M_1{\rm Im}\,\tilde m_{2,3} }{v^2}} = 10^{-6}\frac{{\rm Im}\,\tilde m_{2,3}}{0.05\eV}\frac{M_1}{10^{10}\GeV} 
\end{equation}
where { $\tilde{m}_{2,3}\equiv \lambda_{2,3}^{2} v^2/M_{2,3}$} is
the contribution to the light neutrino mass mediated by the heavy $N_{2,3}$.

The operator argument implies that eq.\eq{epsilon1} holds in any model
where particles much heavier than $M_1$ mediate a neutrino mass operator with
coefficient $\tilde{m}_{2,3}$.
In the past it was debated about if  only the `vertex' diagram in fig.\fig{Ndecay}
or also the `self-energy' diagram
should be included when computing the CP asymmetry: the operator argument makes clear that
both diagrams contribute, since in the limit $M_{2,3}\gg M_1$ the two diagrams
reduce to the same insertion of the $(LH)^2$ operator.

\medskip

The final amount of baryon asymmetry can be written as
\begin{equation}
\label{eq:nB1}
\frac{n_B}{n_\gamma} \approx \frac{ \epsN \eta}{g_{\rm SM}}
\end{equation} 
where $ g_{\rm SM} = 118$ is the number of spin-degrees of freedom of SM particles
(present in the denominator of eq.\eq{nB1}
because only $N_1$ among the many other particles in the thermal bath
generates the asymmetry)
and $\eta $ is an efficiency factor that depends on 
how much out-of-equilibrium $N_1$-decays are.

\subsection{Efficiency}\label{etasec}
We now discuss this issue.
If $N_1\to LH $ decays  are slow enough,
the $N_1$ abundancy does not decrease according to the Boltzmann equilibrium statistics 
$n_{N_1}\propto e^{-M_1/T}$ demanded by thermal equilibrium,
so that late out-of-equilibrium $N_1$ decays generate a lepton asymmetry.
Slow enough decay means  $N_1$ lifetime longer than the inverse expansion rate.
At $T\sim M_1$ one has
$$ R \equiv {\Gamma_1\over H(M_1)}\sim {\tilde{m}_1\over \tilde{m}^*}\qquad\hbox{where}\qquad
\tilde{m}^*\equiv \frac{256\sqrt{g_{\rm SM}} v^2}{3M_{\rm Pl}}=2.3~10^{-3}\eV$$
is fixed by cosmology.
All the dependence on the mass and Yukawa couplings of $N_1$ is incorporated in 
$\tilde{m}_1\equiv \lambda^2_1 v^2/M_1$,
the contribution to the light neutrino mass mediated by $N_1$.
Unfortunately $\tilde{m}_1$ and $\tilde{m}_{2,3}$ are only related to the observed atmospheric and solar mass
splittings in a model-dependent way.
Unless neutrinos are almost degenerate
(and unless there are cancellations)
$\tilde{m}_1$ and $\tilde{m}_{2,3}$ are smaller than $m_{\rm atm}\approx 0.05\eV$.
\medskip

\begin{figure}[t]
$$\includegraphics[width=\textwidth]{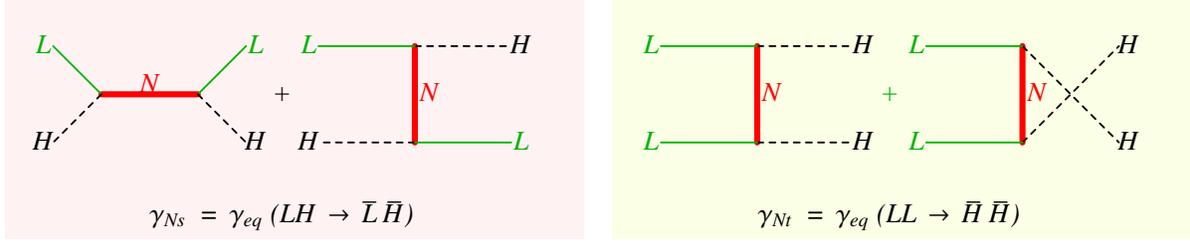}$$
\caption{\em Wash-out $LH \leftrightarrow \bar{L}\bar H$ and $LL \leftrightarrow \bar H \bar H$
$\Delta L = 2$ scatterings.
\label{fig:Nscatt2}}
\end{figure}

If $R \ll 1$ (i.e.\ $N_1$ decays strongly out-of-equilibrium) then $\eta=1$.

If instead $R \gg 1$ the lepton asymmetry is only mildly suppressed as $\eta\sim1/R$.
The reason is that  $N_1$ inverse-decays,
which tend to maintain thermal equilibrium by regenerating
decayed $N_1$,
have rates suppressed by a Boltzmann factor at $T<M_1$:
$R(T < M_1)\approx R \cdot  e^{-M_1/T} $. 
The $N_1$ quanta that decay when $R(T)<1$, i.e.\ at $T< M_1/\ln R$,
give rise to unwashed leptonic asymmetry.
At this stage the $N_1$ abundancy is suppressed by the Boltzmann factor $e^{-M_1/T} = 1/R$.
In conclusion, the suppression factor is approximately given by
\beq\label{eq:eff0}
\eta \sim \min(1,\tilde{m}^*/\tilde{m}_1)\qquad
\hbox{(if $N_1$ initially have thermal abundancy)}.\eeq
Furthermore, we have to take into account that 
virtual exchange of $N_{1,2,3}$ gives rise to
$\Delta L=2$ scatterings  (see fig.\fig{Nscatt2}) that wash-out the lepton asymmetry.
Their thermally-averaged interaction rates are relevant only at
$M_1\circa{>}10^{14}\GeV$, when $N_{1,2,3}$ have large ${\cal O}(1)$ Yukawa couplings.
When relevant, these scatterings give a strong exponential suppression of the baryon
symmetry, because 
their rates are not suppressed at $T\circa{<} M_1$ by
a Boltzmann factor (no massive $N_1$ needs to be produced).

\medskip

\begin{figure}[t]
$$\includegraphics[width=0.45\textwidth]{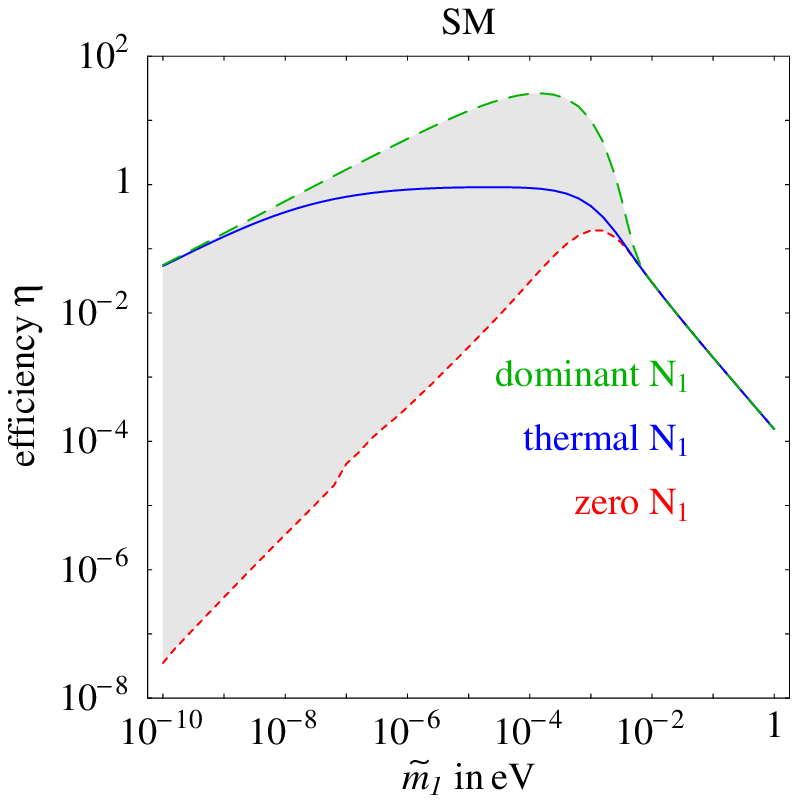}\hspace{0.1\textwidth}
\includegraphics[width=0.45\textwidth]{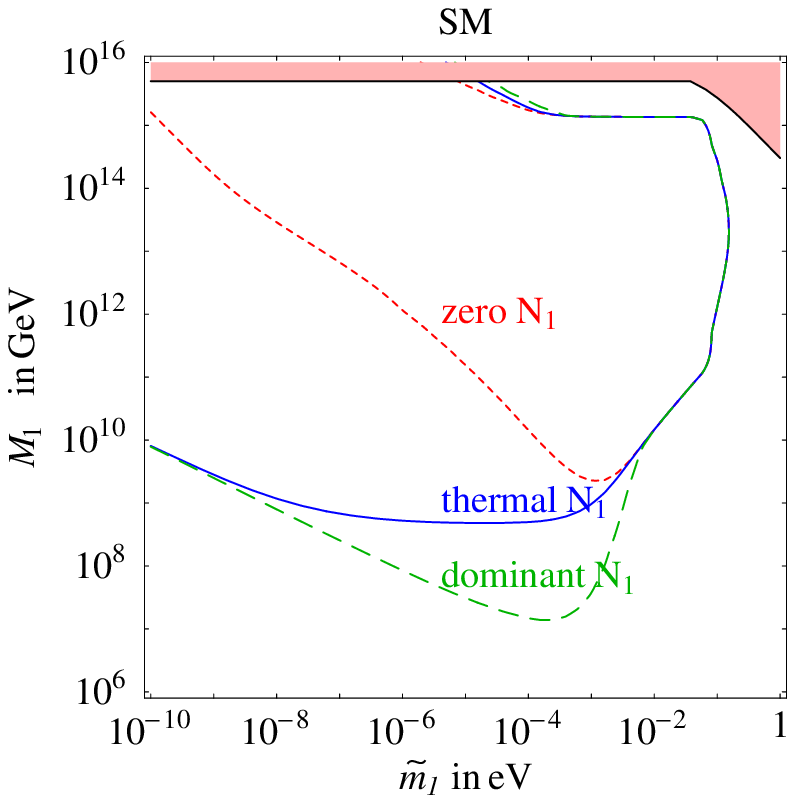}$$
\caption[x]{\em 
Fig.\fig{leptogEta}a: efficiency $\eta$ as function of 
$\tilde{m}_1$ for $M_1=10^{10}\GeV$
and for different assumptions about the initial $N_1$ abundancy;
for $\tilde{m}_1$ larger than a few {\rm meV} the efficiency is univocally predicted to be
$\eta(\tilde{m}_1) \approx 0.42 ({\rm meV}/\tilde{m}_1)^{1.15}$.
Fig.\fig{leptogEta}b: the regions in the $(\tilde{m}_1,M_1)$ plane
inside the curves can lead to successful leptogenesis.
\label{fig:leptogEta}}
\end{figure}

So far we assumed that right-handed neutrinos have thermal initial  abundancy.
Let us discuss how the result depends on this assumption.
If $\tilde{m}_1\gg \tilde{m}^*$ (in particular if $\tilde{m}_1=m_{\rm atm}$ or
$m_{\rm sun}$) 
the efficiency does not depend on the assumed initial conditions,
because decays and inverse-decays bring the $N_1$ abundancy close to thermal equilibrium.
For $\tilde{m}_1\circa{<} \tilde{m}^*$ the result depends on the unknown initial condition.
\begin{itemize}
\item If $N_1$ have negligible initial abundancy at $T\gg M_1$ 
and are generated only by the processes previously discussed,
their abundancy at $T\sim M_1$ is suppressed by $\tilde{m}_1/\tilde{m}^*$.
Therefore the efficiency factor is approximatively given by
\beq\label{eq:eta0}\eta \sim \min(\tilde{m}_1/\tilde{m}^*,\tilde{m}^*/\tilde{m}_1)\qquad
\hbox{(if $N_1$ initially have zero  abundancy)}.\eeq
\item Finally, in the opposite limit where $N_1$ initially dominate the energy density of the universe,
the suppression factor $1/g_{\rm SM}$ in eq.\eq{nB1} no longer applies, and
the efficiency factor can reach $\eta\sim g_{\rm SM}$.
\beq\label{eq:etainf}\eta \sim \min(g_{\rm SM},\tilde{m}^*/\tilde{m}_1)\qquad
\hbox{(if $N_1$ initially have dominant  abundancy)}.\eeq
\end{itemize}
These estimates agree with the results of a detailed numerical
computation, shown in fig.\fig{leptogEta}a.
Notice that if $\tilde{m}_1\gg  \tilde{m}^*$ the $N_1$ abundancy gets close enough
to thermal equilibrium, such that the lepton asymmetry generated
by $N_1$ decays does not depend on the initial $N_1$ abundancy.
Furthermore, $N_1$ decays and inverse-decays typically wash-out
a possible pre-existing lepton asymmetry. 
 
In the next section we describe how a precise computation can be done.
This is not necessary for understanding the  final discussion of section~\ref{final}.

\section{Thermal leptogenesis:  precise computation}\label{Boltz}
As previously discussed many important computations in cosmology are done
using Boltzmann equations. So it is useful to have this tool.

\subsection{Boltzmann equations}\label{Boltz2}
In absence of interactions the number of particles in a comoving volume $V$ remains constant.
Boltzmann equations allow to follow the effect of different interactions.
Let us study e.g.\ how $1\leftrightarrow 2+3$ decay and inverse-decay processes
affect the number $n_1$ of `1' particles
(in the case of leptogenesis we have $N_1\leftrightarrow LH$):
\begin{eqnarray}
\label{eq:Boltz1} \frac{d}{dt}(n_1 V) &=&V \int d\vec{p}_1\int d\vec{p}_2\int d\vec{p}_3\,
(2\pi)^4 \delta^4(p_1-p_2-p_3)\times\\
&&\times\nonumber [-|A_{1\to 23}|^2 f_1(1\pm f_2)(1\pm f_3) + |A_{23\to 1}|^2(1\pm f_1) f_2 f_3]\end{eqnarray}
where $d\vec{p}_i = d^3p_i/2E_i(2\pi)^3$ is the relativistic phase space,
$|A_{1\to 23}|^2$ and $|A_{23\to 1}|^2$
are the squared transition amplitude summed over initial and final spins,
and $f_i$ are the energy (and eventually spin, flavour, color,...) distributions of the various particles.
To start we assume that CP violation can be neglected, such that the direct and the inverse
process have a common amplitude $A$.

In line of principle we should study the evolution of all $f$'s in order to obtain the total densities
$ n = \sum \int f~d^3p/(2\pi)^3$.
In practice elastic scatterings (i.e.\ interactions that do not change the number of particles)
are typically fast enough that they maintain {\em kinetic equilibrium},
so that the full Boltzmann equations for $f$ are solved by 
$f(p) = f_{\rm eq}(p){n}/{n_{\rm eq}}$ where 
  $ \displaystyle  f_{\rm eq} = [e^{E/T}\pm 1]^{-1}$ are the Bose-Einstein and Fermi-Dirac distributions.
Each particle species is simply characterized by its total
abundancy $n$,
that can be varied only by  inelastic processes.

  The factors $1\pm f_i$ in eq.\eq{Boltz1} take into account Pauli-Blocking (for fermions) and stimulated emission (for bosons).
Since the average energy is $\langle E\rangle \sim 3T$ within $10\%$ accuracy one can approximate
with the Boltzmann distribution
$f_{\rm eq}\approx e^{-E/T}$ and set $1\pm f \approx 1$.
This is a significant simplification.

When inelastic processes are sufficiently fast to maintain also {\em
chemical equilibrium},
the total number $n_{\rm eq}$ 
of particles with mass $M$ at temperature $T$ are
$$
n_{\rm eq}=g\!\int {d^3p~f _{\rm eq}\over(2\pi \hbar)^3}=\frac{gM^2T}{2\pi^2}
{\rm K}_2(\frac{M}{T}) = \left\{\begin{array}{ll}
{gT^3}/{\pi^2} & T\gg M\\
g(MT/2\pi)^{3/2}e^{-M/T}& T\ll M
\end{array}\right.
$$
where $g$ is the number of spin, gauge, etc degrees of freedom.
A right handed neutrino has $g_N = 2$, a photon has $g_\gamma=2$,
the 8 gluons have $g_{G^a} = 16$, and all SM particles have $g_{\rm SM} = 118$.
The factor $\hbar=h/2\pi$ has been explicitly shown to clarify the physical origin of the 
$2\pi$ in the denominator.

The Boltzmann equation for $n_1$ simplifies to
\begin{eqnarray} \frac{1}{V}\frac{d}{dt}(n_1 V) &=& \int d\vec{p}_1\int d\vec{p}_2\int d\vec{p}_3\,
(2\pi)^4 \delta^4(p_1-p_2-p_3)\times\\
\nonumber &&\times  |A|^2[- \frac{n_1}{n_1^{\rm eq}}
e^{-E_1/T} + \frac{n_2}{n_2^{\rm eq}}
\frac{n_3}{n_3^{\rm eq}}e^{-E_2/T}e^{-E_3/T}]
\end{eqnarray}
One can recognize that the integrals over final-state momenta reconstruct the decay rate $\Gamma_1$,
and that the integral over $d^3p_1/E_1$ averages it over the thermal distribution of initial state particles; the factor $1/E_1$ corresponds to Lorentz dilatation of their life-time.
Therefore the final result is
\beq\label{eq:Boltz2} \frac{1}{V}\frac{d}{dt}(n_1 V)
= \langle \Gamma_1 \rangle  n_1^{\rm eq}\left[ \frac{n_1}{n_1^{\rm eq}}- \frac{n_2}{n_2^{\rm eq}}
 \frac{n_3}{n_3^{\rm eq}}\right] 
\eeq
where $\langle \Gamma_1\rangle$ is the thermal average of the Lorentz-dilatated decay width 
\beq\Gamma_1(E_1) =\frac{1}{2E_1} \int d\vec{p}_2\, d\vec{p}_3 (2\pi)^4\delta^4(p_1-p_2-p_3) |A|^2.\eeq
Analogous results holds for scattering processes.

 If $\langle \Gamma_1\rangle \gg H$ the term in square brackets in eq.\eq{Boltz2} is forced to vanish.
 This just means that interactions much faster than the expansion rate force chemical equilibrium,
giving    $\mb{n} = \mb{n}_{\rm eq}$.

 In the case of leptogenesis $2,3=L,H$ have fast gauge interactions.
 Therefore we do not have to write and solve Boltzmann equations for $L,H$, because
 we already know their solution: $L,H$ are kept in equilibrium.
 We only need to insert this result in the Boltzmann equation for $N_1$, that simplifies to
\beq \dot{n}_1 +3H n_1 = \langle\Gamma_1\rangle (n_1 - n_1^{\rm eq})\eeq
having used $\dot{V}/V =3H=-\dot s/s$.

In computer codes one prefers to avoid very big or very small numbers:
it is convenient to reabsorb the $3H$ term (that  accounts for the dilution due to the overall
expansion of the universe) by normalizing the number density $n$ to
the entropy density $s$.
Therefore we study the evolution of  $Y = n/s$,
as function of $z=m_N/T$
in place of time $t$
($H\, dt = d\ln R = d\ln z$ since during adiabatic
expansion $sV$ is 
constant, i.e.\ $V\propto 1/T^3$).

\medskip

Using $Y(z)$ as variables, the general form of Boltzmann equations is
 \beq \hspace{-2mm} sHz\frac{dY_1}{dz} = \sum
 \Delta_{1}\cdot
 \gamma_{\rm eq}(12\cdots \leftrightarrow 34\cdots)\left[
 \frac{Y_1}{Y_1^{\rm eq}}\frac{Y_2}{Y_2^{\rm eq}}\cdots-
 \frac{Y_3}{Y_3^{\rm eq}}\frac{Y_4}{Y_4^{\rm eq}}\cdots\right]\eeq
where the sum runs over all processes that vary the number of
`1' particles by $\Delta_1$ units (e.g.\ $\Delta_1 = -1$ for $1\to 23$ decay,
$\Delta_1 = -2$ for $11\to 23$ scatterings, etc.)
and $\gamma_{\rm eq}$ is the specetime density (i.e.\ the number per unit volume and unit time)
in thermal equilibrium of the various processes.

Neglecting CP-violating effects, direct and inverse processes 
have the same reaction densities.
For a scattering and for its inverse process one gets the previous result:
$$ \gamma_{\rm eq}(1\,\to \,23) = 
\int d\vec{p}_1 \, f_1^{\rm eq}   \int  d\vec{p}_2\,d\vec{p}_3  
\, (2\pi)^4\delta^4(p_1-p_2-p_3) |A|^2  = \gamma_{\rm eq}(23\,{\to} \,1)$$
The thermal average of the decay rate can be analytically computed in terms of Bessel functions:
\beq\label{eq:gammaD} \gamma_{\rm eq}(1\to 23\cdots)=\gamma_{\rm eq}(23\cdots\to
1)=
      n^{\rm eq}_1
{\mbox{K}_1(z)\over\mbox{K}_2(z)}\,\Gamma(1\to 23\cdots) \eeq

For a 2-body scattering process
$$ \gamma_{\rm eq}(12\,{\leftrightarrow}\,34) = 
\int d\vec{p}_1\, d\vec{p}_2 \, f_1^{\rm eq}  f_2^{\rm eq}  \int  d\vec{p}_3\,d\vec{p}_4
\, (2\pi)^4\delta^4(p_1+p_2-p_3-p_4) |A|^2  $$
When there are $n$ identical particles in the final or initial states one should divide
by a $n!$ symmetry factor.
One can analytically do almost all integrals, and obtain
$$  \gamma_{\rm eq}(12\to 34\cdots)={T\over 32\pi^4}\int_{s_{\rm min}}^{\infty}
ds~s^{3/2}\, \lambda(1,M_1^2/s,M_2^2/s)  {\sigma}(s)\  
\mbox{K}_1\!\left(\frac{\sqrt{s}}{T}\right) 
$$
which is the thermal average of $v\cdot \sigma$, summed over initial and final state spins.

\begin{figure}
$$\includegraphics[width=\textwidth]{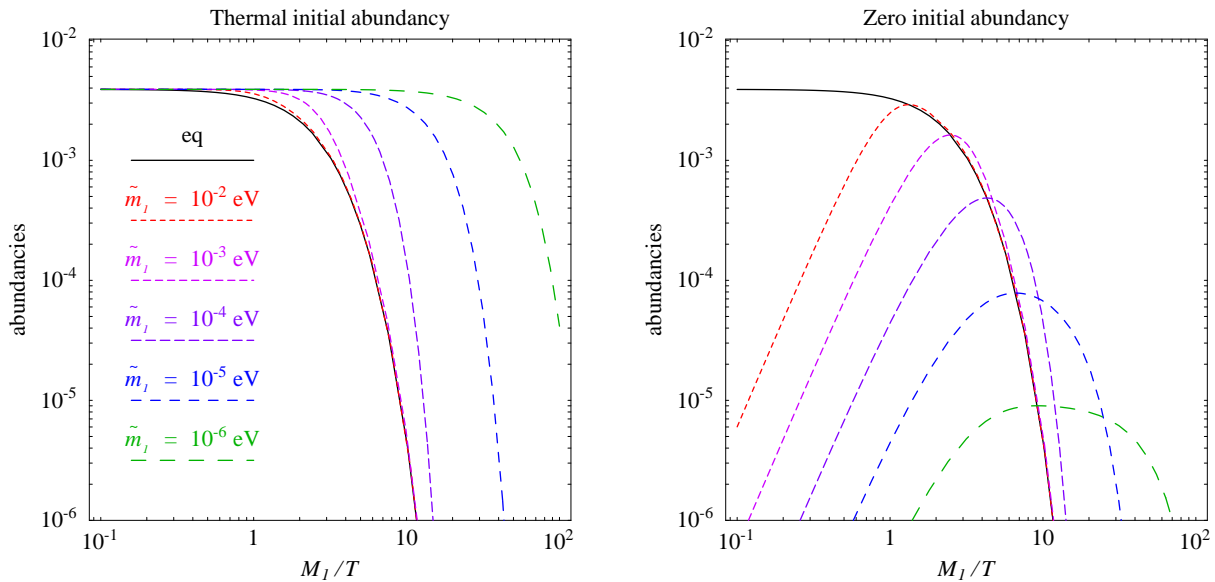}$$
 \caption{\label{fig:Ysol}\em Evolution of $Y_{N_1}(M_1/T)$ for different values of $\tilde{m}_1$ starting from
 different initial conditions. The continuous line is the abundancy in thermal equilibrium. }
\end{figure}

\subsection{Boltzmann equations for leptogenesis}
We now specialize to leptogenesis.
The main process is $N_1\leftrightarrow  HL,\bar H\bar L$ decay and inverse decay.
These processes are enough to generate a $N_1$ population that follows 
the Boltzmann distribution\footnote{Intuitively,
one would probably guess a different, incorrect, result: that $HL\to N_1$ inverse decays
generate $N_1$ with the average energy of two particles, rather than with the average energy of one particle, since $E_{N_1}=E_L+E_H$,
and that two particles have more energy than one.
However, using $e^{-E_L/T}\cdot e^{-E_H/T}= e^{-E_{N_1}/T}$ one verifies that
$\langle E_{N_1}\rangle = \langle E_L+E_H\rangle $: this is why 
Boltzmann found that the thermal distribution is exponential.},
so that we can write the Boltzmann equation for the total $N_1$ abundancy.
We denote with $\gamma_D$ its equilibrium density rate,
computed inserting
$\Gamma(N_1) =  \lambda_1^2 M_1/8\pi$ in eq.\eq{gammaD}.
The Boltzmann equation for the $N_1$ abundancy is
\beq \label{eq:N1eq} 
sHz \frac{dY_{N_1}}{dz} = -\gamma_D (\frac{Y_{N_1}}{Y_{N_1}^{\rm eq}}-1).\eeq
Fig.\fig{Ysol} shows howe $Y_{N_1}$ evolves for different values of $\tilde{m}_1$.
As expected if $\tilde{m}_1\gg 10^{-3}\eV$ one gets a result close to thermal equilibrium
independently of the initial condition.

In order to get the Boltzmann equation for the lepton asymmetry one needs to take
into account the small CP-violating terms.
Let us start by including only the $\Delta L=1$
CP-violating $N_1\to HL,\bar{H}\bar{L}$ decays.
We write the decay rates in terms of the CP-conserving total 
decay rate $\gamma_D$ and of the CP-asymmetry $\epsN\ll1$: 
\begin{equation}\label{eq:gammaD2}
\begin{array}{l}
\gamma_{\rm eq}(N_1\to LH)=\gamma_{\rm eq}(\bar{L}\bar{H}\to N_1)=
(1+\epsN){\gamma_D}/{2},\\
\gamma_{\rm eq}(N_1\to \bar{L}\bar{H})=\gamma_{\rm eq}(LH\to N_1)=
(1-\epsN){\gamma_D}/{2} .
\end{array}\end{equation}
In this approximation the number of leptons $L$ and anti-leptons $\bar L$ evolve as
\bea 
zHs Y'_L &=&\frac{\gamma_D}{2}\bigg[\frac{Y_{N_1}}{Y_{N_1}^{\rm eq}} 
(1+\epsN)-\frac{Y_L}{Y_L^{\rm eq}} (1-\epsN)\bigg],\\
zHs Y'_{\bar{L}} &=& \frac{\gamma_D}{2}\bigg[\frac{Y_{N_1}}{Y_{N_1}^{\rm eq}} 
(1-\epsN)-\frac{Y_{\bar L}}{Y_{\bar L}^{\rm eq}} (1+\epsN)
\bigg].\eea
Here $Y_{N_1}^{\rm eq}$, $Y_L^{\rm eq}$ and $Y_{\bar L}^{\rm eq}$ are
equilibrium densities each with 2 degrees of freedom.
Ignoring ${\cal O}(\epsN^2)$ terms, the lepton asymmetry $\AL = Y_L - Y_{\bar L}$
evolves as
\beq sHz \AL' =  \epsN \gamma_D(\frac{Y_N}{Y_N^{\rm eq}} ~{+}~1)-
\frac{\AL}{2 Y_L^{\rm eq}}\gamma_D.\eeq
The second term describes how  $\gamma_D$
tends to restore thermal equilibrium, washing out the lepton asymmetry.
{\em The first term makes no sense}: it would generate a lepton asymmetry even in thermal
equilibrium, $Y_{N_1} = Y_{N_1}^{\rm eq}$, violating Sakharov conditions.
An acceptable Boltzmann equation would contain
 ${Y_N}/{Y_N^{\rm eq}} ~{-}~1$, but we made no sign error.
Indeed, taking into account only decays and inverse decays, an asymmetry 
is really generated even in thermal
equilibrium, since CPT invariance implies
that if $N_1$ decays preferentially produce $L$, 
than inverse decays preferentially destroy $\bar L$ i.e.\
they have the same net effect.

\subsubsection*{A subtlety: avoiding overcounting}
To obtain correct Boltzmann equations one must include all processes that contribute
at the chosen order in the couplings.
The CP-asymmetry is generated at  ${\cal O}(\lambda^4)$: at this order
we must include also the $\Delta L=2$ scatterings of fig.\fig{Nscatt2}; we name their rates as
\beq \gamma_{Ns} \equiv \gamma_{\rm eq}(LH \leftrightarrow\bar{L}\bar{H}),\qquad
\gamma_{Nt}\equiv\gamma_{\rm eq}(LL \leftrightarrow\bar{H}\bar{H}),\eeq
and $\gamma_{\Delta L = 2}\equiv 2( \gamma_{Ns}+\gamma_{Nt})$.
At first sight it is enough to include these scatterings at tree level,
obtaining CP-conserving reaction densities $\gamma={\cal O}( \lambda^4)$
that cannot correct our non-sensical CP-violating term.
This is basically right, although the true argument is more subtle.

Indeed $LH \leftrightarrow\bar{L}\bar{H}$ can be mediated by on-shell $N_1$ exchange
(see fig.\fig{Nscatt2}a):
as usual in these situations (e.g.\ the $Z$-peak) resonant enhancement gives 
$\sigma_{\rm peak}\propto \lambda^0$ in an energy range $\Delta E \propto \lambda^2$,
so that $\gamma_{Ns} \propto\sigma_{\rm peak}\cdot \Delta E\propto  \lambda^2$.
(We will soon obtain the exact result).
Nevertheless one can proof that the reaction density
remains CP-conserving up to one-loop order: unitarity demands
$\sum_j |M(i\to j)|^2 = \sum_j |M(j\to i)|^2$, so
$$\sigma(LH\to LH) + \sigma(LH\to\bar L\bar H) = \sigma(LH\to LH) + \sigma(\bar L\bar H \to LH)$$
(at higher order states with more particles allow a negligible CP asymmetry).

\medskip


%

The key subtlety is that  the $LH\leftrightarrow \bar L\bar H$ 
scattering rate mediated by $s$-channel exchange of $N_1$
shown in fig.\fig{Nscatt2}a,
must be computed by subtracting the CP-violating contribution 
due to on-shell $N_1$ exchange,
because in the Boltzmann equations this effect is already taken 
into account by successive decays,
$LH\leftrightarrow N_1\leftrightarrow\bar L\bar H$.
Since the on-shell contribution is 
$$\gamma_{Ns}^{\rm on-shell}(
LH\to\bar{L}\bar{H}) =\gamma_{\rm eq}(LH\to N_1) \hbox{BR}(N_1\to \bar{L}\bar{H}),$$
where ${\rm BR}(N_1\to \bar{L}\bar{H})=(1-\epsN)/2$, we obtain 
\begin{eqnarray}
\gamma_{\rm eq}^{\rm sub}(LH\to\bar{L}\bar{H})&=&\gamma_{Ns}-(1-\epsN)^2
{\gamma_D}/{4},\\
\gamma_{\rm eq}^{\rm sub}(\bar{L}\bar{H}\to LH)&=&\gamma_{Ns}-(1+\epsN)^2
{\gamma_D}/{4},
\end{eqnarray}
Including subtracted scatterings  at leading order in $\epsN$ gives the final Boltzmann equation~\cite{leptogenesis}
\begin{eqnarray}
zHs \AL' &=&\gamma_D \bigg[\epsN 
\bigg(\frac{Y_{N_1}}{Y_{N_1}^{\rm eq}}-1\bigg) - 
\frac{\AL}{2Y_{L}^{\rm eq}}\bigg]\label{eq:B2}
 -  2\gamma_{\Delta L=2}^{\rm sub}  \frac{\AL}{Y_{L}^{\rm eq}} ,
\end{eqnarray}
where $\gamma_{\Delta L=2}^{\rm sub} = \gamma_{\Delta L=2} - \gamma_D/4$.
 
Equivalently, one can more simply not
include the decay contribution $\gamma_D$ to washout of $\AL$
 because it is already accounted by resonant $\Delta L=2$ scatterings.
 Then one gets
\beq \label{eq:B3}
zHs \AL' =\gamma_D\epsN \left(\frac{Y_{N_1}}
{Y_{N_1}^{\rm eq}}-1\right) -2\gamma_{\Delta L=2}\frac{\AL}{Y_{L}^{\rm eq}} , 
\eeq
 which is equivalent to\eq{B2}.
 Using eq.\eq{B3} it is not necessary to compute subtracted rates.
 (Subtraction is performed incorrectly in works before 2003).

\subsubsection*{Including sphalerons and Yukawas}
Finally, we have to include sphaleronic scatterings, 
that transmit the asymmetry from left-handed leptons 
to left-handed quarks, generating a baryon asymmetry.
Similarly we have to include SM Yukawa couplings, that transmit the asymmetry to
right-handed quarks and leptons.

In theory one should enlarge Boltzmann equations adding all these processes.  
In practice, depending on the value of $M_1$, during leptogenesis at $T\sim M_1$ 
these process often give reaction densities which are either
negligibly slower or much faster than the expansion rate:
in the first case they can be simply neglected, 
in the second case they simply enforce thermal equilibrium.
One can proceed by converting the Boltzmann equation for $\AL$ into 
a Boltzmann equation for $\YBL $:
since $\YBL $ is not affected by these redistributor processes
we only need to find how processes
in thermal equilibrium relate $\YBL $ to $\YL $.
Sphalerons and the  $\lambda_{t,b,c,\tau}$ Yukawas are fast at $T\circa{<}10^{11\div12}\GeV$.
At larger temperatures all redistribution processes are negligibly slow and one trivially has
$\YBL= - \YL $.
At intermediate temperatures one has to care about flavor issues, discussed later.

It is interesting to explicitly compute redistribution factors at $T\sim\TeV$ when all redistributor processes are fast.
Each particle $P=\{L,E,Q,U,D,H\}$ carries an asymmetry $A_P$.
Interactions equilibrate `chemical potentials' $\mu_P\equiv A_P/g_P$ as
$$\left\{\begin{array}{ll}
ELH\hbox{ Yukawa}: & 0=\mu_E+\mu_L+\mu_H\\
DQH\hbox{ Yukawa}: &0=\mu_D+\mu_Q+\mu_H\\
UQ\bar{H}\hbox{ Yukawa}: &0=\mu_U+\mu_Q-\mu_H\\
QQQL\hbox{ sphalerons}:&0=3\mu_Q+\mu_L\\
\hbox{No electric charge}: &0=N_{\rm gen}(\mu_Q-2\mu_U+\mu_D-\mu_L+\mu_E)-2N_{\rm Higgs}\mu_H
\end{array}\right.$$
Solving the system of 5 equations and 6 unknowns, 
one can express all asymmetries in terms of one of them, conveniently chosen to be $\YBL$:
\begin{eqnarray*}
\YB &=&N_{\rm gen}(2\mu_Q-\mu_U-\mu_D)=\frac{28}{79}(\YBL),\\
\YL &=&\YB-(\YBL)=-\frac{51}{79}(\YBL).
\end{eqnarray*}
The efficiency $\eta$ is precisely defined such that the final baryon asymmetry is 
\beq\label{eq:eta}\frac{n_B}{s}= \YB=  -\frac{28}{79} \epsilon \eta Y_{N_1}^{\rm eq}(T\gg M_1)\qquad\hbox{i.e.}\qquad
\left.\frac{n_B}{n_\gamma}\right|_{\rm today} = -\frac{\epsN\eta}{103.}\eeq
in agreement with the estimate\eq{nB1}.

\medskip

Various extra processes give corrections of relative order $g^2/\pi^2,\lambda_t^2/\pi^2\sim \hbox{few}~\%$. 
Some of these corrections have been already computed:
scattering involving gauge bosons and/or top quarks.
Others have not yet been included: three body $N_1$-decays,
one-loop correction to the $N_1\to LH $ decay and its CP-asymmetry.
Thermal corrections have not been fully included.
The fact that $\gamma_D$ is the only really relevant rate makes a full 
inclusion of these subleading corrections feasible.

\subsubsection*{Including flavor}
So far we studied the dynamics of leptogenesis in `one-flavor' approximation, eq.\eq{Lseesaw}.
In the literature, flavor was fully included only after that  neutrino data showed that
flavor mixing among leptons can be large.
The one-flavor approximation remains useful because
including flavor adds so many unknown parameters that a precise discussion is impractical:
e.g.\
we do not know which combination of $L_e$, $L_\mu$, $L_\tau$ is the lepton doublet
$L$ coupled to $N_1$ in the see-saw Lagangian, eq.\eq{Lseesaw}.
To include flavor,
the Bolztmann equation for $Y_{{\cal L}}$ must be generalized to an evolution
equation for the $3\times 3$ matrix density $\rho$ of lepton asymmetries in each flavor. 
However, it simplifies to qualitatively different behaviors in different ranges of $M_1$:

\begin{itemize}
\item If $M_1\circa{>}10^{11}\GeV$ all SM lepton Yukawa couplings induce scattering
rates much slower than the expansion rate $H$ at $T\sim M_1$:
quantum coherence among different flavors survives undamped and
the main new effect is that lepton asymmetries generated by $N_2,N_3$ decays 
can be washed-out by processes involving $N_1$ 
only along the combination of flavors to which $N_1$ couples.
So, one must sum the contributions of all right-handed neutrinos produced
after reheating.

\item 
If $M_1\circa{<}10^{9}\GeV$, the $\lambda_\mu$ and $\lambda_\tau$ Yukawa couplings induce scattering rates faster than  $H$ at $T\sim M_1$
  and damp quantum coherence in $\rho$: the matrix equation for $\rho$ reduces to 3
  Boltzmann equations for the asymmetries in the $\ell=\{e,\mu,\tau\}$ flavors.
  Neglecting a mild flavor mixing (induced by sphaleronic scatterings)
    these equations have the following form: 
  eq.\eq{N1eq} for $Y_{N_1}$ remains unchanged, and eq.\eq{B3} for $\AL$
 splits into three equations for $\AL_e$, $\AL_\mu$, $\AL_\tau$ with the CP-violating term and the wash-out term restricted to each flavor,
  as intuitively expected.
  Therefore, in each flavor $\ell$ one has a different CP asymmetry $\varepsilon_{1\ell}$
  and  efficiency
  $\eta_\ell(\tilde{m}_1,\tilde{m}_{1\ell})$ 
 that now depends on two parameters: the usual flavor independent 
 $\tilde{m}_1$ that tells the total $N_1$ decay rate, 
 and the flavor-dependent $\tilde{m}_{1e,\mu,\tau}\equiv |\lambda_{1\ell}^2| v^2/M_1$ 
 that parameterize wash-out.
  A good approximation is~\cite{leptogenesis}: 
  
 \noindent
\parbox{0.6\textwidth}{\beq
\eta_\ell(\tilde{m}_1,\tilde{m}_{1\ell})\approx \eta(\tilde{m}_{1\ell}).\eeq
where
$\eta$ is the one-flavor efficiency, plotted in fig.\fig{leptogEta}a.
 The side figure shows the exact numerical result of $\eta_\ell$
 as function of $\tilde{m}_{1\ell}$ for
 different values of  $\tilde{m}_1/\tilde{m}_{1\ell}=1$
 (continuous line), 2 (red dotted line), 4 (blue dashed line), confirming 
 that the result negligibly depends on $\tilde{m}_1$, especially if it is
 close to the values suggested by solar or atmospheric oscillations.
 Therefore eq.\eq{nB1} gets replaced by
 \beq\label{eq:flavoredapprox}
   n_B/n_\gamma \approx  \sum_{\ell} \varepsilon_{1\ell} \cdot \eta(\tilde{m}_{1\ell})/g_{\rm SM}.
\eeq  
 For large mixing angles one typically has $ \tilde{m}_1/\tilde{m}_{1\ell}\sim {\cal O}(2)$:
 the above approximation tells that taking flavor into account
 can enhance the efficiency by an ${\cal O}(2)$ factor. }
\parbox{0.4\textwidth}{\hspace{0.02\textwidth}\includegraphics[width=0.3\textwidth]{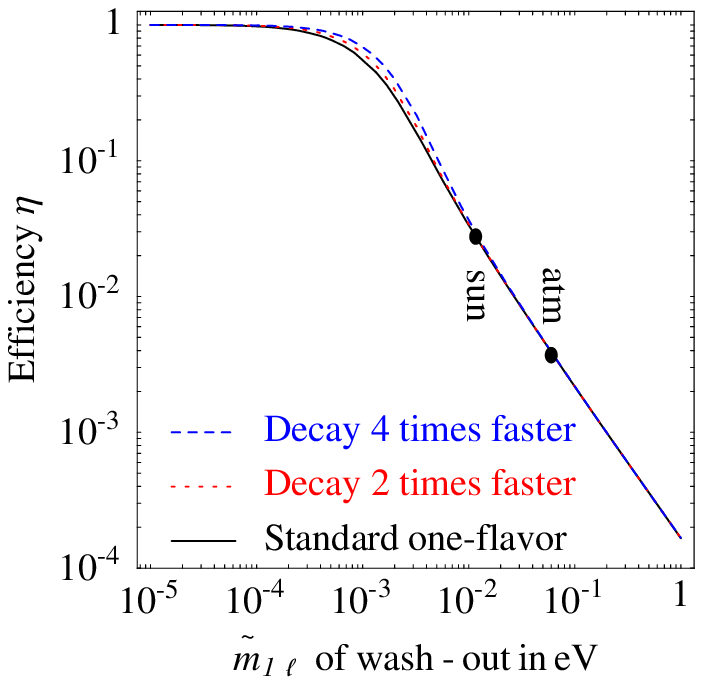}}

\item Something intermediate happens if $10^9\GeV\circa{<}M_1\circa{<}10^{11}\GeV$:
quantum coherence stays undamped only among $\mu$ and $e$, such that one must take into
account the asymmetry possibly generated by $N_{2},N_3$ along $e,\mu$.

\end{itemize}

\section{Testing leptogenesis?}\label{final}
Unfortunately speculating that neutrino masses and the baryon asymmetry
are produced by the see-saw mechanism and by thermal leptogenesis
is much easier than testing them.


A direct test seems impossible, because
right-handed neutrinos are either too heavy or too weakly coupled
to be produced in  accelerators.

What about indirect tests?
We trust BBN because it predicts the primordial abundances of several nuclei in terms
of known particle physics.
Leptogenesis explains a single number, $n_B/n_\gamma$, in terms of speculative physics at
high energies: the see-saw model has 18 free parameters.
Neutrino masses only allow to measure 9 combinations
of these parameters, and thereby provide a too weak link.
The situation might improve if future experiments will confirm
certain supersymmetric models, in which
quantum corrections imprint neutrino Yukawa couplings $\lambda_{ij}N_i L_jH$
in slepton masses, inducing lepton flavour violating (LFV) processes such as  $\mu\to e\gamma$, $\tau\to\mu\gamma$,
$\tau\to e\gamma$ with possibly detectable rates.
Measuring them, in absence of other sources of LFV, would roughly allow us to measure the 3
off-diagonal entries of $\lambda^\dagger\cdot\lambda$.
Detectable LFV rates are obtained if $\lambda\circa{>}10^{-(1\div2)}$.
In any case, reconstructing all see-saw parameters in this way is unrealistic.
Maybe future experiments will discover supersymmetry, LFV in charged leptons,
and will confirm that neutrino masses violate lepton number and CP,
and we will be able to convincingly argue that this can be considered as circumstantial
evidence for see-saw and thermal leptogenesis.
Archeology is not an exact science.

\smallskip

Another possibility is that we might find a correctly predictive model of flavour.
Presently  three approaches give some predictions:
symmetries, numerology, zerology.
Symmetries can be used to enforce relations like  
$\theta_{23} = \pi/4$, $\tan^2\theta_{12}=1/2$, $\theta_{13}=0$
where $\theta_{ij}$ are the neutrino mixing angles.
Numerology can suggest relations like $\theta_{12} + \theta_{C} = \pi/4$.
Zerology consists in assuming that 
flavour matrices have many  negligibly small entries;
 for example one can write see-saw textures with only one CP-violating phase.
 This scheme does not allow to predict the sign of CP-violation in neutrino oscillations 
in terms of the sign of the baryon asymmetry,
because the sign of the baryon asymmetry also depends on which right-handed neutrino is the lightest one.

\subsection{Constraints from leptogenesis}
We here discuss testable constraints.
Although this topic is tortuous, we avoid over-simplifications,
at the price of obtaining a tortuous section.

As discussed in section~\ref{lepto}, 
assuming that $N_1$ is lighter enough  than other sources of neutrino masses
that their effects can be fully encoded in the neutrino mass operator,
the CP-asymmetry $\epsN$ is directly connected with  it.
Under the above hypothesis, one can derive constraints from leptogenesis.
To do so, we need to generalize eq.\eq{epsilon1} taking flavour into account.
Denoting flavour matrices with boldface, 
we define $\tilde{\mb{m}}_i$ to be the contribution to the neutrino mass matrix mediated
by the right-handed neutrino $N_i$, so that
$\mb{m}_\nu = \tilde{\mb{m}}_1+\tilde{\mb{m}}_{2}+\tilde{\mb{m}}_{3} $.
Then
\beq\label{eq:DI}  | \epsN|  = \frac{3}{16\pi}\frac{M_1}{v^2}\frac{|{\rm Im}\,{\rm Tr} \,\tilde{\mb{m}}_1^\dagger 
(\tilde{\mb{m}}_{2}+\tilde{\mb{m}}_{3})|}{\tilde{m}_1}
\le
 \frac{3}{16\pi}\frac{M_1}{v^2}(m_{\nu_3} - m_{\nu_1})\eeq
where 
$m_{\nu_3}$ ($m_{\nu_1}$) denotes the mass of the heaviest (lightest) neutrino.
Rather than rigorously proofing the constraint\eq{DI}~\cite{leptogenesisBounds}, let us understand its origin and limitations in a simpler way.

\begin{itemize}
\item[1)] 
Let us start considering the case of hierarchical neutrinos: $m_{\nu_1}\ll m_{\nu_3}$:
the constraint is obtained by substituting 
$\tilde{\mb{m}}_{2}+\tilde{\mb{m}}_{3} = \mb{m}_\nu - \tilde{\mb{m}}_1$,
and holds whatever new physics produces $\tilde{\mb m}_{2,3}$.
Since $|\varepsilon_1|\propto M_1$,
one can derive a lower bound on the mass $M_1$ of the lightest right-handed neutrino by
combining eq.\eq{DI} with a precise computation of thermal leptogenesis in one-flavor approximation
and with the measured  baryon asymmetry and neutrino masses:
\beq\label{eq:mNbound}
M_1 >  \left\{\begin{array}{rl}
2.4\times 10^{9} \GeV    &   \hbox{if $N_1$ has zero} \\
 4.9\times 10^{8}\GeV     &\hbox{if $N_1$ has thermal} \\  
1.7\times 10^{7}\GeV     &\hbox{if $N_1$ has dominant}
\end{array}\right.
\hbox{initial abundancy}
\eeq
 and assuming $M_{1}\ll M_{2,3}$.
 Including flavor as discussed previously relaxes the first constraint by a factor ${\cal O}(2)$.

\item[2)] 
The factor $m_{\nu_3} - m_{\nu_1}$ is specific to the see-saw model with 3 right-handed neutrinos.
To understand it, let us notice that 3 right-handed neutrinos can produce
the limiting case of degenerate neutrinos $m_{\nu_1} = m_{\nu_2}=m_{\nu_3}=m_\nu$
only in the following way:
each $N_i$  gives mass $m_\nu$ to one neutrino mass eigenstate.
Since they are orthogonal in flavour space,
the CP-asymmetry of eq.\eq{DI} vanishes due to flavour orthogonality:
this is the origin of the $m_{\nu_3} - m_{\nu_1}$ suppression factor.

The bound\eq{DI} implies an upper bound on the mass of quasi-degenerate neutrinos:
$m_\nu \circa{<}0.2\eV$~\cite{leptogenesisBounds}.
Indeed for large $m_\nu$ both the efficiency and  the maximal CP-asymmetry become smaller,
because heavy neutrinos must be quasi-degenerate, and $\tilde{m}_1 \ge m_{\nu_1}$,
$m_{\nu_3}- m_{\nu_1} \simeq \Delta m^2_{\rm atm}/2m_{\nu}$.
Furthermore, the bound can be improved up to 
$m_{\nu_3} < 0.15\eV$~\cite{leptogenesisBounds} ($3\sigma$ confidence level) by computing the upper bound on $| \epsN|$
 for given $\tilde{m}_1$
and maximizing $n_B$ with respect to $\tilde{m}_1$
taking into account how the efficiency of thermal leptogenesis decreases for large $\tilde{m}_1$.
The leptogenesis constraint is very close to observed neutrino masses, and stronger than 
experimental bounds.
\end{itemize}

\medskip

However, this leptogenesis constraint holds under the 
dubious assumption 
that hierarchical right-handed neutrinos produce quasi-degenerate neutrinos, 
while good taste suggests that quasi-degenerate
neutrinos are more naturally  produced by quasi-degenerate
right-handed neutrinos.
In general the constraint\eq{DI} evaporates if 
the particles that mediate $\tilde{\mb{m}}_{23}$
are so light that their effects cannot be encoded in $\tilde{\mb{m}}_{23}$:
the CP-asymmetries becomes sensitive to the detailed structure of
the neutrino mass model.
Suppression due to flavor-orthogonality was the most delicate consequence of our 
initial assumption, and is the
first result that disappears when they are relaxed, allowing
$M_{2,3}$ to be not much heavier than $M_1$.
Correspondingly, the constraint on quasi-degenerate neutrino masses,
that heavily relies on the factor $m_{\nu_3}-m_{\nu_1}$,
becomes weaker than experimental constraints.

\smallskip

Furthermore, 
in the extreme situation where right-handed neutrinos are very degenerate,
$M_2 - M_1 \circa{<}\Gamma_{1,2}$
a qualitatively new effect appears:
CP violation in $N_1\leftrightarrow N_2$ mixing.
This phenomenon is fully analogous to  $K^0 \leftrightarrow  \bar K^0$ mixing, and
for $M_2 - M_1 \sim\Gamma_{1,2}$ it allows 
a maximal CP-asymmetry, $|\epsN|\sim 1$.
This means that with a tiny $M_2-M_1$ 
 one can  have successful thermal leptogenesis even at the weak scale.

\medskip

The constraint\eq{mNbound} is more robust,
but we still have to clarify what the assumption $M_{2,3}\gg M_1$ means in practice.
Surely one needs $M_2-M_1\gg \Gamma_{1,2}$ such that 
only CP-violation in $N_1$ decay is relevant.
The issue is: is $M_{2,3}/M_1\sim 10$ 
(a hierarchy stronger than the one present in left-handed neutrinos)
hierarchical enough to guarantee that the constraint holds, up to $1/10^2=\%$ corrections?
The answer is no: an operator analysis allows to understand 
how the constraint can be completely relaxed. 
The physics that above $M_1$ produces the dimension-5 neutrino mass operator 
$(LH)^2/2$
can also produce
a related dimension-7 operator $\Upsilon\equiv (LH)\partial^2 (LH)/2$,
that contributes to $\epsN$.
Since neutrino masses do not constrain $\Upsilon$, it can be large enough
to over-compensate the suppression due to its higher dimension.
We make the argument more explicit, by considering the concrete case of
see-saw models, where above $M_1$ there are two other 
right-handed neutrinos with masses $M_2$ and $M_3$.
Including `dimension-7' terms suppressed by $M_1^2/M_{2,3}^2$ and dropping inessential
${\cal O}(1)$ and flavour factors,
the CP-asymmetry becomes:
\beq\epsN \sim \frac{3}{16\pi}\frac{M_1}{v^2}
{\rm Im} \bigg[ \tilde{m}_2 ( 1 + \frac{M_1^2}{M_2^2}) + \tilde{m}_3 (1 + \frac{M_1^2}{M_3^2})\bigg].\eeq
At leading order $\epsN$ depends only on $\tilde{m}_2+\tilde{m}_3$, that,
as previously discussed, cannot be large and complex.
At higher order $\epsN$ depends separately on $\tilde{m}_2$ and $\tilde{m}_3$,
that can be large and complex provided that their sum stays small.
One can build models where this naturally happens, obtaining an
$\epsN$ orders of magnitudes above the DI bound.
The enhancement is limited only by perturbativity, $\lambda_{2,3}\circa{<}4\pi$.
In supersymmetric models, large Yukawa couplings lead to the testable effects previously discussed.

\subsection{Leptogenesis and supersymmetry}
Adding supersymmetry affects some ${\cal O}(1)$ factors.
Eq.\eq{eta} remains almost unchanged, because adding spartners roughly doubles
both the number of particles that produce the baryon asymmetry and
the number of particles that share it.
Ignoring small supersymmetry-breaking terms, right-handed neutrinos and sneutrinos
have equal masses, equal decay rates and equal CP-asymmetries.
Both $\Gamma_{N_1}$ and $\epsN$ become 2 times larger,
because there are new decay channels.
As a consequence of more CP-asymmetry compensated by more wash-out,
the constraints on right-handed and left-handed neutrino masses
discussed in the non-supersymmetric case remain essentially unchanged.

\medskip

The leptogenesis constraint\eq{mNbound} on $M_1$ acquires a new important impact:
in many supersymmetric models  the maximal temperature at which the Big-Bang started 
(or, more precisely, the `reheating temperature' $T_{\rm RH}$) must be less than about 
$T_{\rm RH}\circa{<}10^7\GeV$,
in potential conflict with the constraint\eq{mNbound}.

Let us discuss the origin of the supersymmetric constraint on $T_{\rm RH}$.
Gravitinos are the supersymmetric partner of the graviton:
they have a mass presumably
not much heavier than other sparticles,  $m_{\tilde{G}}\circa{<}\TeV$,
and gravitational couplings to SM particles.
Therefore gravitinos decay slowly after BBN (life-time 
$\tau \sim M_{\rm Pl}^2/m_{\tilde{G}}^3\sim {\rm sec} \,(100\TeV/m_{\tilde{G}})^3$):
their decay products damage the nuclei generated by BBN.
The resulting bound on the gravitino abundancy depends on unknown gravitino branching
ratios.\footnote{Gravitinos might be stable if they are the lightest supersymmetric particle.
But in this case dangerous effects are produced by gravitational decays of the next-to-lightest SUSY particle
into gravitinos.}
The gravitino interaction rate is $\gamma_{\tilde G}(T)\sim T^6/M_{\rm Pl}^2$, which means that
gravitinos have been  generated around the reheating temperature $T_{\rm RH}$,
with abundancy $n_{\tilde G}/n_\gamma \sim \gamma_{\tilde G}/Hn_\gamma \sim T_{\rm RH}/M_{\rm Pl}$.
Therefore gravitinos suggest an upper bound on $T$.

This is why in the previous section we carefully discussed
specific scenarios that allow
low-scale thermal leptogenesis.
Supersymmetry suggests a new scenario named `soft leptogenesis':
complex soft terms in the see-saw sector
give new contributions to the CP-asymmetry,
$\varepsilon_1 \sim \alpha_2 m_{\rm SUSY}^2/M_1^2$,
which can be significant if $M_1$ is not much larger than the SUSY-breaking scale $m_{\rm SUSY}$
(presumed to be below 1 TeV).
At larger $M_1$ `soft leptogenesis' can still be relevant,
but only in a fine-tuned range of parameters.


\medskip

\begin{figure}
$$\includegraphics[width=\textwidth]{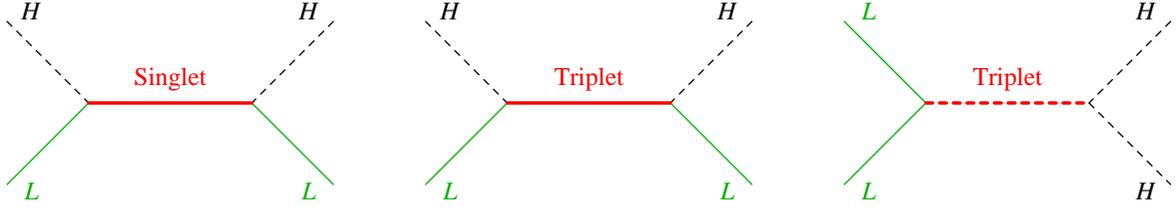}$$
\caption{\em 
The Majorana neutrino mass operator $(LH)^2$ can be mediated by tree level exchange of: {\rm I)} a fermion singlet; {\rm II)} a fermion  triplet;
{\rm III)} a scalar triplet.\label{fig:123}}
\end{figure}

\subsection{Leptogenesis in alternative neutrino-mass models}
Generic neutrino masses can be mediated by tree-level exchange of
three different kinds of new particles~\cite{seesaw}:
I) at least three fermion singlets;
II) at least three fermion SU(2)$_L$ triplets; 
III) one scalar  SU(2)$_L$ triplet
(or of combinations of the above possibilities).
Fig.\fig{123} shows the relevant Feynman diagrams.

So far we studied case I). 
Can leptogenesis be used to distinguish between these possibilities?
Leptogenesis can be produced in decays of $P$, the lightest particle
that mediates neutrino masses.
The neutrino-mass contribution to its CP-asymmetry  is 
given by expressions similar to eq.\eq{DI},  with
$\tilde{\mb{m}}_1$ generalized to be the contribution to
neutrino masses mediated by $P$,
and $\tilde{\mb{m}}_{2}+\tilde{\mb{m}}_{3}$ generalized to be
the contribution of all heavier sources.

\medskip

It was expected that the efficiency $\eta$ can be high enough
only if $P$ is a right-handed neutrino:
as discussed in section~\ref{etasec} it easily decays out-of-equilibrium giving
\beq
\eta(\hbox{fermion singlet})\approx \min\left[X,\frac{H}{\Gamma}\right]\eeq
where $\Gamma$ is its decay rate, $H$ is the expansion rate at $T\sim M$ and
\beq X=\left\{\begin{array}{ll}
1&\hbox{for thermal}\\
\Gamma/H&\hbox{for negligible}\\
g_{\rm SM}&\hbox{for dominant}
\end{array}\right.\hbox{initial abundancy}.\eeq
A  $\SU(2)_L$ triplet (scalar or fermion)
has gauge interactions that
keep its abundancy  close to thermal equilibrium so that
the 3rd Sakharov condition cannot be fulfilled.
This suppression is present, and a quantitative analysis is needed to
see if/when it is strong enough.
The Boltzmann equation for the triplet abundancy $Y$ has an extra term $\gamma_A$
that accounts for annihilations of two triplets into gauge bosons:
\beq sHz \frac{dY}{dz} = -\gamma_D (\frac{Y}{Y_{\rm eq}}-1)-2
\gamma_A (\frac{Y^2}{Y^2_{\rm eq}}-1).\eeq
The term $\gamma_A \sim g^4 T^4$ is dominant only at $T\circa{>} M$,
where $M$ is the triplet mass;
at lower temperatures it is strongly suppressed by a double Boltzmann factor $(e^{-M/T})^2$,
because gauge scattering must produce 2 triplets.
The resulting efficiency can be approximated as
\beq
\eta(\hbox{fermion triplet})\approx \min\left[1,\frac{H}{\Gamma},\frac{M}{10^{12}\GeV}\max(1,\frac{\Gamma}{H})\right].
\eeq
 $\eta$ is univocally predicted, because at $T\gg M$ 
gauge interactions thermalize the triplet abundancy.
At $T\sim M_1$ gauge scatterings partially annihilate triplets:
in section~\ref{DM} we learnt how to estimate how many particles
survive to annihilations, and this is the origin of
the factor $M/10^{12}\GeV$ in $\eta$.
The last factor takes into account that
annihilations are ineffective if triplets decay before annihilating.

As illustrated in fig.\fig{123}
a scalar triplet separately decay to leptons, with width $\Gamma_L(T^*\to LL)$,
and in Higgses, with width $\Gamma_H(T\to HH)$.
Lepton number is effectively violated only when both processes are faster
than the expansion of the universe, giving
\beq
\eta(\hbox{scalar triplet})\!\approx\! \min\!\left[1,\frac{H}{\min(\Gamma_L,\!\Gamma_H)},\frac{M}{10^{12}\GeV}\max(1,\frac{\Gamma_L\!+\!\Gamma_H}{H})\right].
\eeq
This means that a quasi-maximal efficiency $\eta\sim 1$ is obtained when
 $T^*\to LL$ is faster than gauge annihilations
while $H\to TT$ is slower than the expansion rate.
In conclusion, leptogenesis from decays of a $\SU(2)_L$ triplet can be sufficiently efficient
even if triplets are light enough to be tested at coming accelerators, $M\sim\TeV$.


\subsection{Leptogenesis and dark matter in loop-mediated neutrino-mass models}
Neutrino masses mediated at one-loop level can be realized
by exchanging various possible kinds of new particles;
some of them are potential DM candidates.
For example, one can introduce the usual right-handed neutrinos $N$ and couple them as
$\lambda\, NLH'$, where $H'$ is not the usual Higgs scalar doublet, but another scalar doublet coupled to the Higgs
as $\lambda' (H^\dagger H')^2+\hbox{h.c}$.
The Lagrangian is restricted to couplings invariant under the Z$_2$ symmetry
$N\to - N$ and $H'\to - H'$, such that the lightest component of
these particles is stable and is a  DM candidate.

Neutrino masses are suppressed by a one loop factor $r \sim \lambda'/(4\pi)^2$ with respect to the standard see-saw.
Leptogenesis depends on the coupling $\lambda$ and on $M_{1,2,3}$ in the usual way:
therefore, at fixed values of neutrino masses, 
the CP asymmetry ($\propto \tilde{m}_{2,3}$, see eq.\eq{epsilon1})
is enhanced by $1/r$ with respect to the standard case,
while the efficiency factor $\eta$
(that depends on $\tilde{m}_1$ as in eq.s~(\ref{eq:eff0}--\ref{eq:etainf}))
is suppressed by $r$ if $\tilde{m}_1\sim m_{\rm sun,atm}$.
The constraints in eq.\eq{mNbound} on $M_1$ are relaxed by the factor $1/r$.

\medskip

DM might be generated in the usual way discussed in section~\ref{DM}.
Alternatively, a new interesting possibility is that 
{\em both DM and leptogenesis might be generated by the same out-of-equilibrium process}.
In the example above, $N_1$ decays generate an asymmetry both in $L$ and $H'$:
$H'$ could be DM with abundancy dominated by its asymmetry (like protons) rather than by the
usual freeze-out relic abundancy discussed in section~\ref{DM}.
{\em If}  the DM asymmetry  is not washed-out, 
this scenario leads to a precise testable prediction for the DM mass:
$m_{\rm DM}= c m_p \Omega_{\rm DM}/\Omega_b\sim 10\GeV$,
where $c$ is a model-dependent ${\cal O}(1)$ factor, analogous to the 28/79 in eq.\eq{eta}.

However the DM asymmetry survives only until
$H$ gets a vev breaking the electroweak gauge symmetry: this happens
at temperatures $T \circa{<} 1.2 m_h$.
Indeed DM is neutral under unbroken electromagnetism
and a Z$_2$-like symmetry guarantees DM stability but does not protect the DM asymmetry.
One can avoid wash-out by promoting Z$_2$ to a global U(1): 
we now discuss why this class of models seems not viable.
In the model used as concrete example, imposing a global U(1) means 
setting $\lambda'=0$, such that
$H'$ is not affected when $H$ develops its vev.
However, direct searches for DM told that DM must not only be
neutral under electromagnetism, but also almost neutral under the $Z$:
the DM $Z$ coupling must be $10\div 100$ times smaller than a typical electroweak coupling~\cite{DM}.
If the DM multiplet (either scalar of fermionic)
lies in a complex representation of the electroweak gauge group
(such that it can carry an asymmetry), reduced $Z$ couplings need  $\lambda'$-like couplings.
E.g., in the concrete  $H'$ example, $\lambda'$ splits 
the complex neutral component of $H'$, $S+iA$, into
a real scalar and pseudo-scalar, $S$ and $A$, with different masses.
The $Z_\mu$ does not couple to the lightest DM mass eigenstate $S$ or $A$
but only to the off-diagonal
combination $A\cdot \partial_\mu S - S\cdot\partial_\mu A$,
without giving unseen DM signals if the mass difference $\Delta m\equiv |m_S-m_A|$ 
is bigger than about $20\,{\rm keV}$
(the expected kinetic energy of DM around the earth).
One can show that the needed $\Delta m$ is so big that
oscillations among $S+iA$ and $S-iA$ destroy the DM asymmetry.\footnote{This needs a non trivial analysis:
Bolztmann equations for the density matrix of $S$ and $A$ show that 
DM-number-conserving gauge scatterings with rate $\Gamma\gg H$
synchronize oscillations among different moment and reduce their
frequency from $\Delta m$ to $\Delta m^2/\Gamma$.
These non-trivial features help a lot, but not enough.}
To avoid this conclusion one needs either
$m_{\rm DM}\approx 30 m_h$ or models where DM is
a complex neutral scalar singlet coupled to the Higgs:
both possibilities look unattractive.

\paragraph{Acknowledgments}
The original work presented in the last section was performed in collaboration with
M. Raidal and V. Rychkov, that agreed on briefly summarizing here our no-go result.
I thank S. Davidson, A. Riotto and T. Kashti for useful discussions.
Since these are lessons,
the bibliography prefers later systematic works 
to pioneering imperfect works.



\small


\begin{thebibliography}{1}


\bibitem{WMAP} 
CMB anisotropies: 
\hepart[astro-ph/0603449]{D.N.~Spergel {\it et al.} (WMAP Science Team)}.
\art[astro-ph/0310725]{SDSS collaboration}{Astrophys. J.}{606}{702}{2004}.
Baryon acoustic peak  in matter inhomogeneities:  
\hepart[astro-ph/0608635]{W.J. Percival et al.} and previous SDSS works.
The physics necessary to understand these measurements
is clearly presented in the book `{\em Modern Cosmology}'
by S. Dodelson.

\bibitem{seesaw} {\bf Models of neutrino masses}.
Neutrino masses from right-handed neutrinos (`see-saw'):
\art{P. Minkowski}{\PL}{B67}{421}{1977}.
Neutrino masses from scalar triplets:
\art{M.~Magg and C.~Wetterich}{\PL}{B94}{61}{1980}.
Neutrino masses from fermion triplets:
\art{R. Foot, H. Lew, X.-G. He, G.C. Joshi}{Z. Phys.}{C44}{441}{1989}.
Neutrino masses at one loop: see \art[hep-ph/9805219]{E. Ma}{\PRL}{81}{1171}{1998}
for a list of possibilities. 


\bibitem{Baryogenesis}
{\bf Baryogenesis}.
Sakharov  conditions:
\art{A.D. Sakharov}{JETP Lett.}{91B}{24}{1967}.
Anomalies:
\art{G. 't Hooft}{\PRL}{37}{37}{1976}.
\art{G. 't Hooft}{\PR}{D14}{3432}{1976}.
Sphalerons:
\art{N.S. Manton}{\PR}{D28}{2019}{1983}.
\art{V. Kuzmin, V.A. Rubakov and M.E. Shaposhnikov}{\PL}{155B}{36}{1985}.
\art{J. Ambjýrn, T. Askgaard, H. Porter and M.E. Shaposhnikov}{\NP}{B353}{346}{1991}.
\art[hep-ph/9609481]{P. Arnold, D. Son and L.G. Yaffe}{\PR}{D55}{6264}{1997}.


\bibitem{leptogenesis} {\bf Thermal leptogenesis}.
 \art{M. Fukugita, T. Yanagida}{\PL}{B174}{45}{1986}.
 
The CP-asymmetry has been computed in
\art[hep-ph/9712468]{E. Roulet, L. Covi, F. Vissani}{\PL}{B424}{101}{1998}. 
See also
\art[hep-ph/9710460 version 2]{W. Buchm\"uller, M. Pl\"umacher}{\PL}{B431}{354}{1998};
\art[hep-ph/9805427]{M. Flanz, E.A. Paschos}{\PR}{D58}{11309}{1998}.

First attempts of solving Boltzmann equations:
 \art{M.A. Luty}{\PR}{D45}{455}{1992};
 \art[hep-ph/9604229]{M. Pl\"umacher}{Z. Phys.}{C74}{549}{1997}.
RGE corrections and flavor were included in
\art[hep-ph/9911315]{R. Barbieri, P. Cre\-minelli, N. Tetradis, A. Strumia}{\NP}{B575}{61}{2000}.
The analytical approximation in eq.\eq{flavoredapprox} to  flavored leptogenesis 
was employed in
\art[hep-ph/0210021]{M. Raidal, A. Strumia}{Phys. Lett.}{B553}{72}{2003}
to precisely study the predictions of a flavor model.
For recent numerical studies of flavored Boltzmann equations see:
\art[hep-ph/0601084]{E. Nardi, Y. Nir, E. Roulet, J. Racker}{JHEP}{01}{164}{2006};
\hepart[hep-ph/0605281]{A. Abada, S. Davidson, F. Josse-Michaux, M. Losada, A. Riotto}.
\hepart[hep-ph/0607330]{S. Blanchet, P. di Bari}.


Thermal  corrections, correct subtraction of resonant processes, reheating effects
have been included in
\art[hep-ph/0310123]{G.F. Giudice, A. Notari, M. Raidal, A. Riotto, A. Strumia}{Nucl. Phys.}{B685}{89}{2004}.

Supersymmetric leptogenesis has been first studied in
\art[hep-ph/9704231]{M. Pl\"umacher}{\NP}{B530}{207}{1998}.
`Resonant leptogenesis' with quasi-degenerate right-handed neutrinos: see e.g.\
\art[hep-ph/9707235]{A. Pilaftsis}{Phys. Rev.}{D56}{5431}{1997}.
\art[hep-ph/0309342]{A. Pilaftsis, T. Underwood}{Nucl. Phys.}{B692}{303}{2004}.
\hepart[hep-ph/0507092]{G.C. Branco et al.}.
`Soft leptogenesis': 
for a systematic treatment see
\art[hep-ph/0407063]{Y.~Grossman, T.~Kashti, Y.~Nir, E.~Roulet}{JHEP}{0411}{080}{2004}.




\bibitem{leptogenesisBounds} {\bf Constraints from leptogenesis}.
The maximal CP asymmetry $\epsN$ of eq.\eq{DI} was rigorously proofed in
\art[hep-ph/0202239]{S. Davidson, A. Ibarra}{Phys. Lett.}{B535}{25}{2002}.
Constraint on neutrino masses: see
\art[hep-ph/0302092]{W.~Buchmuller, P.~Di Bari, M.~Pl\"umacher}{\NP}{B665}{445}{2003}
and 
\art[hep-ph/0312203]{T. Hambye et al.}{Nucl. Phys.}{B695}{169}{2004}.
Mildly hierarchical right-handed neutrinos allow a much larger $\epsN$, as discussed in the previous paper.
Leptogenesis from fermion triplets was studied in the previous paper.
Leptogenesis from scalar triplets: for a systematic treatment see
\hepart[hep-ph/0510008]{T. Hambye, M. Raidal, A. Strumia}.



  \bibitem{DM}
  For a recent review  see
  \art{G. Jungman, M. Kamionkowski, K. Griest}{Phys. Rep.}{267}{195}{1996}.
  Strongest constraint: \hepart[astro-ph/0509259]{CDMS collaboration}.
  

 \end{thebibliography}
\end{document}